\documentstyle[11pt,paspconf]{article}

\begin{document}
\def\gtorder{\mathrel{\raise.3ex\hbox{$>$}\mkern-14mu
                \lower0.6ex\hbox{$\sim$}}}
\def\ltorder{\mathrel{\raise.3ex\hbox{$<$}\mkern-14mu
                \lower0.6ex\hbox{$\sim$}}}
\def\eg{{\it e.g.~}}
\def\etal{et al.~}
\def\asec{^{\prime\prime}}
\title{Host Galaxies of Lensed Luminous Quasars at $\langle z\rangle\sim 2$}

\author{Hans-Walter Rix\altaffilmark{1}, E.~Falco\altaffilmark{3},
C.~Impey\altaffilmark{2}, C.~Kochanek\altaffilmark{3}, 
 J.~Lehar\altaffilmark{3}, B.~McLeod\altaffilmark{3}, J. Mu\~noz\altaffilmark{3}, C.~Peng\altaffilmark{2}
}

\altaffiltext{1}{Max-Planck-Institut fuer Astronomie, Heidelberg} 
\altaffiltext{2}{University of Arizona} 
\altaffiltext{3}{Harvard-Smithsonian Center for Astrophysics}



\begin{abstract}
We present H-band observations of gravitationally lensed  QSO host galaxies obtained with NICMOS on HST as part of the CfA-Arizona-Gravitational-Lens-Survey (CASTLES). The detections are greatly facilitated by the lensing magnification in these systems; we find that most  hosts of radio-quiet QSOs (RQQ) at $z \sim 2$ are of modest luminosity ($L < L_{*}$). They are 2-5 times fainter than the hosts of radio-loud QSOs at the same epoch.

Compared to low redshifts, RQQ hosts at $z \sim 2$ also support higher nuclear luminosities  at given stellar host mass. This suggests that the supermassive black holes at their centers grew faster at early epochs than the stellar body of their surrounding host galaxies.   

\end{abstract}




\section{Gravitational Lensing and QSO Host Galaxies}

The seemingly disjoint topics of strong gravitational lensing and 
of the host galaxies that  surround the luminous QSOs at  high
redshift, enjoy a symbiotic relationship.

\smallskip
\noindent{\sl Hosts constrain lensing:~} Gravitationally lensed radio lobes can provide - through their spatially extended nature - more
constraints on the lensing potential than the two or four point
source images that are found in optically selected or radio-quiet QSOs.
As the majority of high-z QSOs are radio-quiet, the  extended stellar light
of the host galaxies that surround  QSOs, can provide a welcome
subsitute for extended radio emission; the host galaxies form ''optical Einstein rings'' or small
arcs when lensed. Recent examples include PG~1115+080 (Impey \etal 1998), 
MG~1131+0456 (Kochanek \etal 1999), Q~0957+561 (Keeton \etal 1999, {\it in prep.}), and B~1938+666 (King \etal 1998).
Because gravitational lensing preserves the surface brightness of the source, different
images of the same host position in the source plane should only differ in surface brightness through 
differential dust extinction along the various image paths. In this way, optical images
of the lensed hosts could provide a more  sensitive probe of the ISM in distant galaxies
than the point sources alone (\eg Falco \etal 1999).

\smallskip
\noindent{\sl Lensing helps to see host galaxies:~} Gravitational lensing magnifies both the
nuclear QSO emission and the surrounding host galaxy. For the unresolved nucleus this 
magnification simply results in a flux increase without observable changes
in the source morphology; in contrast, the extended host emission is ``stretched away'' 
from the nucleus at a constant surface brightness. If the PSF falls off with radius 
$\propto r^{-3}$, as for HST,  gravitational lensing 
changes the local contrast in favor of the extended light proportional to the square of the
magnification.

\begin{figure}
\plotfiddle{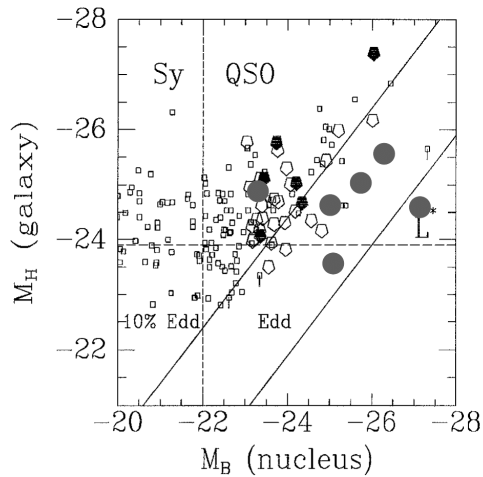}{2.0truein}{0}{35}{35}{-150}{-82}\vskip 0.3truein
\caption{Comparison of the nuclear luminosity to the stellar host luminosity for AGNs. All small symbols represent objects with $z \ltorder 0.5$ (from McLeod \etal 1999). The two diagonal lines represent the expected relations for $M_{BH}/M_{*} \sim 0.005$ (Magorrian \etal 1998) and accrection efficiencies of $ L=0.1 L_{Edd}$ and $L=L_{Edd}$. In contrast, the six large circles represent the results from CASTLES for RQQs at $z \sim 2$. Clearly, these objects have fainter hosts, or more luminous nuclei, than the low-z objects. Indeed, if the H-band M/L were adjusted for the younger populations at $z \sim 2$, most objects studied by CASTLES would have to radiate super-Eddington. } \label{fig-1}
\end{figure}

In the data of the CASTLES  (see {\sl http://cfa-www.harvard.edu/castles/}) project several factors come together 
to provide unprecedented sensitivity to detecting QSO host galaxies: (1) HST provides a relatively stable
PSF with a FWHM of $\sim 0.15\asec$; (2) for a QSO at $z\sim 2$,
 the host galaxy is an order of magnitude more prominent at $1.6\mu m$ compared
to the  AGN than at optical wavelengths; (3) the low near-IR background from space 
improves the sensitivity to low-surface brightness features in the
outer parts of the host galaxy; (4) gravitational lensing further improves the host--AGN contrast
as discussed above. 

For these reasons, we  initiated a study of the host galaxies around
gravitationally lensed quasars. As the lens magnification is physically unrelated to the 
source structure, our sample of lensed QSOs should not be significantly biased compared to other samples, save 
some possible luminosity bias.
In this paper, we summarize briefly the state of other host galaxy studies (\S 2), describe the
results from the CASTLES study (\S 3) and compare them to recent theoretical models
of central black hole and host galaxy growth at high redshifts (\S 4).

\section{Existing Studies of QSO Host Galaxies}

Explicit empirical demonstrations that QSOs are indeed active nuclei surrounded
by host galaxies have been pursued for 25 years, but have been hampered, especially
for distant sources, by contrast problems with the much brighter nucleus, 
particularly because 
 $R_{eff}(host)\approx\sigma_{seeing}\sim1\asec$.

Nonetheless, for QSOs at $z\ltorder 0.5$, host galaxy detections have now become routine
both at visible and near-IR wavelengths (\eg Hutchings and Neff 1992; McLeod and Rieke 1995;
Bahcall \etal 1995), and a fairly coherent picture is emerging: 
radio-quiet QSOs (RQQ) typically live in early-type disk  galaxies (S0-Sb) whose luminosity
range is centered around $L_*$; radio-loud QSOs live in slightly
more luminous  hosts ($L=1-2L_*$). While tidal interactions may be conducive to QSO activity
they are not necessary. McLeod \etal  (1999) showed that in all nearby QSOs ($z<0.5$) there
is a maximum nuclear luminosity for any given H-band host luminosity (Figure 1).
Converting the host luminosity into a stellar mass, $M_*$, and  applying to these QSOs the locally inferred relation $$ M_{BH}/M_*\sim 0.005 $$ (Magorrian \etal 1998) between $M_*$ and the nuclear black hole (BH) mass,
McLeod \etal  (1999) deduce that these QSOs can shine  at most with $L \sim 0.1 L_{Edd}$.

At higher redshifts ($1<z<4$) the host galaxy  picture becomes observationally less clear. 
The hosts of radio-loud QSOs (RLQ)  have been detected in sizeable samples with reasonable completeness 
($\gtorder 50\%$), mostly through near-IR imaging from the ground
(Kotilainen \etal 1998, Carballo \etal 1998). These have become possible
in part because RLQs at these epochs seem to live in extremely luminous galaxies. The observed RLQ hosts
do not seem to have comparably bright present-day counter parts, but it is still not
entirely clear whether most of their (rest) optical emission really is stellar.
In most ground-based searches for RQQ hosts beyond $z=1$ the detection completeness has
been very low: findings are still published on individual objects (\eg Aretxaga \etal 1998).
The relatively few existing detections (\eg Hutchings 1985), and by implication the non-detections,
suggest  that the RQQ hosts are of modest luminosity.

The existing observations leave open a number of important questions about the hosts 
of distant QSOs, which we can address through the lensing studies:

\begin{itemize}
\item Are all QSOs embedded in substantial host galaxies, or are there ``naked'' QSOs?

\item Do the most luminous QSOs (mostly RQQs) at their heyday ($z\sim 2$), 
 live in the most luminous galaxies at their epoch?

\item Do the hosts of RQQs and RLQs differ increasingly with growing redshift?

\item Is there the same maximum nuclear luminosity at a given host luminosity in distant
QSOs than at $z<0.5$? As QSOs were more luminous in the past, and most galaxies were
in lower mass fragments, one might expect an evolution of such a relation.  
\end{itemize}

\begin{figure} 
\plotfiddle{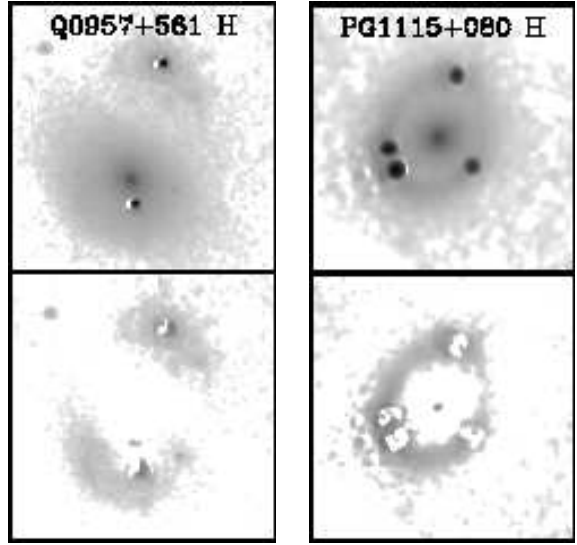}{2.7truein}{0}{80}{80}{-100}{-5}
\caption{Two examples of lensed QSO host galaxies from CASTLES. The top panel shows the original H-band image, containing the multiply imaged nucleus, the lensing galaxy and the host light. Each bottom panel shows the same image after the multiple nucleus images and the lens light have been subtracted, leaving only the host galaxy light.} \label{fig-2}
\end{figure}

\section{CASTLES Observations of RQQ Hosts}

To study gravitationally lensed QSOs systematically, the CASTLES project has obtained deep F160W (H-band) images (typically for one orbit) of most known lens systems. We have begun to derive de-magnified H-band magnitudes for the host galaxies wherever they are detected. In the four image systems we should expect that at least some portion of the host is highly magnified and becomes a detectable arc. Indeed, in most highly magnified cases there is such a qualitative detection, indicating that "naked quasars" are rare, if they exist at all. However, the demagnification of such systems (\eg PG~1115+080, Fig. 2) can be quite complex; certainly, the de-magnification of two-image lenses (\eg Q0957+561, Fig. 2) is more straightforward.

In Figure 3, we have summarized the magnitudes (and upper limit) for an initial set of six objects
(Rix  \etal 1999).
Clearly, these host galaxies of RQQs are much fainter than bright radio galaxies ($m_{H} \approx18$) at similar redshifts.

\begin{figure}
\plotfiddle{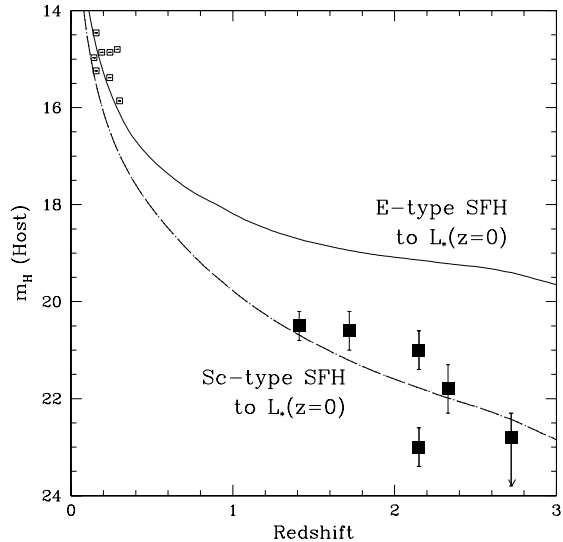}{2.6truein}{0}{38}{38}{-120}{-70}
\caption{Apparent de-magnified fluxes of 6 RQQ hosts detected by CASTLES (solid squares) and low-redshift hosts (open squares) from McLeod and Rieke (1995). For comparison, Figure 3 also shows the apparent brightness of two galaxies with two different star formation histories (SFH) that both result in $L_{*}$ galaxies at the present epoch: for the ``E-type SFH'' all stars are formed in the first
$\sim 10^9$ years, while the ``Sc-type SFH'' has a continuous, slowly declining SFR.} \label{fig-3}
\end{figure}

For comparison, Figure 3 also shows the apparent brightness of two galaxies with two different star formation histories (SFH) that both result in $L_{*}$ galaxies at the present epoch.
If the hosts had formed  all  their stars at high redshift (an $L_{*}$'s worth), they would appear much brighter. However, if the hosts formed stars continuously similar to the Milky Way, the predicted magnitudes match the observations. This is consistent with QSO living in unexceptional host galaxies with typical SFHs that lead to $L_{*}$ now.

\section{Who Grew Faster, Galaxies or their Central Black Holes?}

It is instructive to compare the relation between nuclear and host luminosity for our sources at $z \sim 2$ to the low redshift results of McLeod \etal 1999.

The large filled circles in Figure 1 represent the initial results from CASTLES and show that at higher redshift the hosts are comparatively much fainter than at later epochs. This discrepancy would increase considerably if we compared the minimum stellar mass (rather than H-band luminosity) at high and low redshift.
Indeed, after accounting  for about one magnitude of H-band M/L evolution, we would be lead to conclude that these luminous QSOs at high redshift have super-Eddington luminosities,  {\it if the same $M_{H}/M_{*}$ $\sim 0.005$ relation were to hold.}

Instead, it appears more plausible to infer that the central BHs grew faster at early epochs than the surrounding hosts, so that \newline
$$M_{BH}/M_{*} (z \sim 2) >  M_{BH}/M_{*} (z \leq 0.5). $$ 
In this case, a less massive and less luminous host galaxy could sustain a more luminous nucleus.

Recently, Kauffmann and H\"ahnelt (1999) have explored the connection between galaxy formation and central BH growth through semi-analytic models, trying to explain evolution of the galaxy population and of the QSO luminosity function simultaneously. Their models predict a quite steep evolution in the $L_{Host}$ vs $L_{nucleus}$ relation (see Figure 4), but one that exactly matches our findings.

\begin{figure}
\plotfiddle{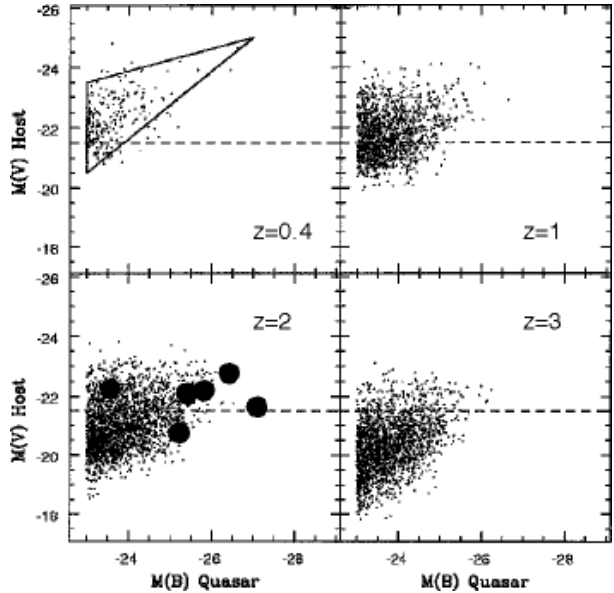}{2.9truein}{-90}{72}{72}{-310}{320}
\caption{Comparison of the host and nuclear luminosity found through CASTLES (large dots) to the predictions from semi-analytic models by Kauffmann and H\"ahnelt (1999) (cloud of small points),
showing the predicted evolution of $M_{V}(Host)$ vs $M_{B}(QSO)$ with redshift. 
The triangle in the top left panel shows the results from McLeod \etal 1999, reiterating
 that our findings from CASTLES are inconsistent with no evolution in $M_{V}(Host)$ vs $M_{B}(QSO)$ from $z < 0.5$ to $z \sim 2$ (see Figure 1). However,  our observed evolution matches well the evolution predicted by Kauffmann and H\"ahnelt.} \label{fig-3}
\end{figure}


\acknowledgments Support for the CASTLES project was provided by NASA through grant numbers GO-7495 and GO-7887 from the Space Telescope Science Institute, which is operated by the Association of Universities for Research in Astronomy, Inc.

%
%

\end{document}